\documentstyle[aps]{revtex}

\begin{document}
\title{{\bf The Continuum Limit and Integral Vacuum Charge}}
\author{{\bf A Calogeracos\thanks{%
Permanent Address, NCA\ Research Associates, PO\ Box 61147, Maroussi 151 01,
Athens, Greece }}}
\author{Low Temperature Laboratory, Helsinki University of Technology,}
\author{PO Box 2200, FIN-02015 HUT, Finland}
\author{{\bf N Dombey}}
\author{Centre for Theoretical Physics, University of Sussex, Brighton, BN1 9QJ, UK}
\maketitle

\begin{abstract}
We investigate a commonly used formula which seems to give non-integral
vacuum charge in the continuum limit. We show that the limit is subtle and
care must be taken to get correct results.
\end{abstract}

\section{Introduction.}

\noindent In this note we consider the continuum limit of a field theory
defined in a large box of size $L$. In the limit $L\rightarrow \infty $, all
discrete states apart from bound states become continuum states. According
to most textbooks, quantum field theory is supposed to be well understood in
this limit. We show, however, that the limiting procedure is subtle and
unless proper care is exercized the usual prescriptions may give incorrect
results. In particular, we show an example where the charge of the system
seems to be a continuous function of the background field whereas it should
be integer-valued.

\smallskip\ 

\noindent An example which we have discussed in previous papers (which are
referred to as CDI \cite{cdi} and CD2 \cite{cd}) is provided by the second
quantized Dirac theory in the presence of a one dimensional four-vector
potential vanishing at spatial infinity. The usual limiting procedure for
counting states in the continuum limit is given by

\begin{equation}  \label{sum}
\sum (states)\rightarrow \frac 1\pi \int^\infty dk\left( L+\frac{d\delta }{dk%
}\right)
\end{equation}

\noindent where $\delta (k)$ is the scattering phase shift for a particle of
momentum $k$ in that limit. Yet as we (and many others) have argued the
vacuum charge $Q_0$ of the system defined in the normal way by the spectral
asymmetry

\begin{equation}  \label{spec}
Q_0=\frac 12\left\{ \sum_k(\text{states with }E>0)-\sum_k(\text{states with }%
E<0)\right\}
\end{equation}

\noindent which is obviously integer-valued in a box of finite size $L$ is
given in the continuum limit using Eq. (\ref{sum}) by

\begin{equation}  \label{err}
Q_0=\frac 12\left\{ \frac 1\pi \left( \delta _{+}(\infty )-\delta
_{+}(0)-\delta _{-}(\infty )+\delta _{-}(0)\right) +N_{+}-N_{-}\right\}
\end{equation}

\noindent where $+$ and $-$ refer to electron and positron scattering phase
shifts and $N_{+}$ and $N_{-}$ are the number of positive and negative
energy bound states. Since in one-dimension (see CDI)

\[
\delta _{\pm }(\infty )=\pm \int_{-\infty }^\infty V(x)dx 
\]

\noindent this implies for the potential $V(x)=\lambda \delta (x)$ that $%
Q_0=\lambda /\pi $ which is a continuous function of $\lambda .$ The correct
result is calculated using Eq. (\ref{spec}) in CD2 to be

\begin{equation}
Q_0=Int\left[ 
{\displaystyle {\lambda \over \pi}}
+%
{\displaystyle {1 \over 2}}
\right]
\end{equation}

\noindent where $Int$ denotes integral part of, which is obviously an
integer. The erroneous result of Eq. (\ref{err}) for the vacuum charge for
this model is found in many places besides CDI, for example in Refs. \cite
{nonint} and \cite{greiner}. The same error may also be responsible for
similar results where other quantised quantities such as baryon charge or
angular momentum seem to be given non-quantised values \cite{vach} for no
apparent reason.

\smallskip\ 

\noindent In CDI following Barton \cite{gb} we counted states and showed
that the number of both positive and negative energy states is unchanged
when a potential is switched on from zero Thus the vacuum charge is still
zero in the presence of a small potential: there is a bound state but the
number of continuum states has decreased by one. As the potential increases
in strength $Q_0$ changes by one according to Eq. (\ref{spec}) whenever a
state crosses $E=0.$ So why is the erroneous result obtained and how can it
be avoided? We now turn to these points.

\smallskip\ 

\noindent In order to quantize the model consistently with standard
anticommutation relations it is essential that eigenfunctions be normalised
to unity. We shall see that to ensure correct normalisation we have to
include an unfamiliar normalisation factor $N$ which reduces to unity as $%
L\rightarrow \infty $; in fact $N^2=1+\tilde O(1/L)$. It turns out that the
expression for $N^2$ involves the phase shifts in a way reminiscent of the
incorrect result of Eq. (\ref{err}). We obtain a new expression for the
charge density differing from the conventional one by terms of order $\sim
1/L$. We find a finite change in the vacuum charge (defined as the space
integral of the vacuum charge density) induced by the spectral asymmetry
residing in a certain region of space of the system. It is crucial to get
this $1/L$ behaviour correct if we wish to sum over states and convert sums
to integrals via 
\begin{equation}  \label{d2}
\sum_k\rightarrow 
{\displaystyle {L \over \pi}}
\displaystyle \int 
_0^\infty dk
\end{equation}

\noindent There are topologically nontrivial models where the Dirac particle
is coupled to a soliton; the attending zero modes induce non-integer values
for the vacuum charge \cite{gold}. However in problems with trivial topology
such as ours $Q_0$ is integral.

\section{The model.}

\subsection{The Normal Modes.}

\noindent The potential $V(z)$ is symmetric and is taken to vanish for $%
\left| z\right| >a.$ We take the system in a box of length $2L$ with
periodic boundary conditions $\psi (-L)=\psi (L).$ Wavefunctions are
classified according to the magnitude of the wavevector outside the well and
parity. Positive energy solutions outside the potential take the form 
\begin{equation}  \label{ee}
u_{e,k}(z)=%
{\displaystyle {N_{e+}(k) \over \sqrt{L}}}
\sqrt{%
{\displaystyle {E+m \over 2E}}
}\left( 
\begin{array}{c}
\cos (kz\pm \Delta _{e+}) \\ 
0 \\ 
{\displaystyle {ik \over E+m}}
\sin (kz\pm \Delta _{e+}) \\ 
0
\end{array}
\right)
\end{equation}
\begin{equation}  \label{eo}
u_{o,k}(z)=%
{\displaystyle {N_{o+}(k) \over \sqrt{L}}}
\sqrt{%
{\displaystyle {E+m \over 2E}}
}\left( 
\begin{array}{c}
i\sin (kz\pm \Delta _{o+}) \\ 
0 \\ 
{\displaystyle {k \over E+m}}
\cos (kz\pm \Delta _{o+}) \\ 
0
\end{array}
\right)
\end{equation}

\noindent The subscript + in the phase shifts $\Delta $ refers to the energy
sign. Similar expressions are valid for negative energy states $v_{e,k}$ , $%
v_{o,k}$ provided we replace $E$ by $\left| E\right| $ and change notation
from $\Delta _{e+},$ $\Delta _{o+},$ $N_{e+}(k),$ $N_{o+}(k)$ to $\Delta
_{e-},$ $\Delta _{o-},$ $N_{e-}(k),$ $N_{o-}(k)$. We also quote for future
reference the form of the even bound state wavefunction outside the well 
\begin{equation}
u_b(\left| z\right| >a)=C\left( 
\begin{array}{c}
1 \\ 
0 \\ 
i%
{\displaystyle {m-E_b \over \kappa}}
\\ 
0
\end{array}
\right) e^{-\kappa z}\text{ , }\kappa =\sqrt{m^2-E_b^2}  \label{bound}
\end{equation}

\noindent We insist on the normalization

\begin{equation}  \label{n1}
\displaystyle \int 
_{-L}^Ldz\psi _k^{\dagger }(z)\psi _k(z)=1
\end{equation}

\noindent for all eigenstates of the Hamiltonian. For the bound state the
appropriate value of $C$ to ensure correct normalization depends on the
detailed behaviour of the potential.

\subsection{The Landau-Lifshitz-Stone lemma.}

We rederive a result originally due to Stone \cite{stone} which itself is
based on a problem in Landau and Lifshitz \cite{ll}. We start with the Dirac
equation; the argument is equally valid for either positive or negative
energy solutions, thus $E_k=\pm \sqrt{k^2+m^2}$%
\begin{equation}
\frac 1i\alpha _z\frac{du_{k^{\prime }}}{dz}+m\beta u_{k^{\prime
}}=E_{k^{\prime }}u_{k^{\prime }}+V(z)u_{k^{\prime }}  \label{e1}
\end{equation}

\noindent and left-multiply by $u_k^{\dagger }$ 
\begin{equation}  \label{e2}
\frac 1iu_k^{\dagger }\alpha _z\frac{du_{k^{\prime }}}{dz}+mu_k^{\dagger
}\beta u_{k^{\prime }}=E_{k^{\prime }}u_k^{\dagger }u_{k^{\prime
}}+V(z)u_k^{\dagger }u_{k^{\prime }}
\end{equation}

\noindent Write the Dirac equation for $u_k$, take the Hermitian conjugate
and right-multiply by $u_{k^{\prime }}$%
\begin{equation}  \label{e3}
-\frac 1i\frac{du_k^{\dagger }}{dz}\alpha _zu_{k^{\prime }}+mu_k^{\dagger
}\beta u_{k^{\prime }}=E_ku_k^{\dagger }u_{k^{\prime }}+V(z)u_k^{\dagger
}u_{k^{\prime }}
\end{equation}

\noindent Subtract (\ref{e3}) from (\ref{e2}) to get 
\begin{equation}  \label{e4}
\frac 1i\frac d{dz}\left( u_k^{\dagger }\alpha _zu_{k^{\prime }}\right)
=\left( E_{k^{\prime }}-E_k\right) u_k^{\dagger }u_{k^{\prime }}
\end{equation}

\noindent Integrating over $z$ from $z_{1\text{ }}$to $z_2$: 
\begin{equation}  \label{e5}
\frac 1i\left[ u_k^{\dagger }\alpha _zu_{k^{\prime }}\right]
_{z_1}^{z_2}=\left( E_{k^{\prime }}-E_k\right) 
\displaystyle \int 
_{z_1}^{z_2}u_k^{\dagger }u_{k^{\prime }}dz
\end{equation}

\noindent Take $k^{\prime }=k+dk$ in the above equation and divide by $dk$: 
\[
\frac 1i\left[ u_k^{\dagger }\alpha _z\frac{du_k}{dk}\right] _{z_1}^{z_2}=%
\frac{dE}{dk}%
\displaystyle \int 
_{z_1}^{z_2}u_k^{\dagger }u_kdz 
\]

\noindent or

\begin{equation}  \label{e6}
\frac 1i\left[ u_k^{\dagger }\alpha _z\frac{du_k}{dk}\right] _{z_1}^{z_2}=%
\frac kE%
\displaystyle \int 
_{z_1}^{z_2}u_k^{\dagger }u_kdz
\end{equation}

\noindent This is the key equation. Its power lies in the fact that to
evaluate the left hand side for $\left| z_1\right| ,\left| z_2\right| >a$ it
suffices to use the asymptotic expressions Eqs. (\ref{ee}, \ref{eo}) for the
wavefunctions where only the phase shifts appear.

\subsection{Normalization of eigenfunctions.}

\noindent Note that the normalisation condition (\ref{n1}) together with
periodic boundary conditions entail restrictions on $k$. Apply relation (\ref
{e6}) at the endpoints $z_1=-L,z_2=L$. Then in the left hand side we only
need the asymptotic expressions Eqs. (\ref{ee}, \ref{eo}) and on the right
hand side we can use Eq.(\ref{n1}) to set the integral equal to unity. The
evaluation of the left hand side simplifies because of the periodic boundary
conditions. We thus obtain 
\begin{equation}
N_{e,o\pm }(k)=\frac 1{\sqrt{1+%
{\displaystyle {1 \over L}}
{\displaystyle {d\Delta _{e,o\pm } \over dk}}
}}  \label{enoe}
\end{equation}

\noindent and therefore 
\begin{equation}  \label{nsq}
N_{e,o\pm }^2(k)=\frac 1{1+%
{\displaystyle {1 \over L}}
{\displaystyle {d\Delta _{e,o\pm } \over dk}}
}\simeq 1-%
{\displaystyle {1 \over L}}
{\displaystyle {d\Delta _{e,o\pm } \over dk}}
\end{equation}

\noindent for $L$ large. Note that the quantities $N_{e,o\pm }^2(k)-1$
vanish both when $V=0$ and in the limit $L\rightarrow \infty $. Eq. (\ref
{enoe}) is the main result of this note.\ 

\section{The Vacuum Charge Density.}

We focus on the charge density 
\begin{equation}  \label{d3}
\rho _k(z)=\psi _k^{\dagger }(z)\psi _k(z)
\end{equation}

\noindent of an eigenstate $\psi _k(z)$ (of definite parity and sign of
energy) corresponding to a particular wavevector $k.$ $\rho _k(z)$ can be
written down for the scattering states for $\left| z\right| >a$ by using the
asymptotic forms (\ref{ee}) and (\ref{eo}) of the wavefunctions. The factors 
$N_{e,o\pm }$ provide a spatially uniform contribution to the charge density
equal to 
\begin{equation}  \label{cont}
-%
{\displaystyle {1 \over L}}
{\displaystyle {d\Delta _{e,o\pm } \over dk}}
\end{equation}

\noindent The effect of Eq. (\ref{cont}) is to modify the charge density in
the presence of the potential. To illustrate this consider the charge $%
Q_{k,ext}$ outside the well due to this state 
\begin{equation}
Q_{k,ext}=\left( 
\displaystyle \int 
_{-\infty }^{-a}+%
\displaystyle \int 
_a^\infty \right) dz\rho _k(z)=2%
\displaystyle \int 
_a^\infty dz\rho _k(z)  \label{d1}
\end{equation}

\noindent since $V$ is symmetric. To order $1/L$ 
\begin{equation}  \label{d5}
Q_{k,ext,even,\pm }=1-%
{\displaystyle {a \over L}}
-%
{\displaystyle {m \over 2LEk}}
\sin 2\left( ka+\Delta _{e\pm }(k)\right) -%
{\displaystyle {1 \over L}}
{\displaystyle {d\Delta _{e\pm } \over dk}}
\end{equation}
\begin{equation}  \label{d7}
Q_{k,ext,odd,\pm }=1-%
{\displaystyle {a \over L}}
+%
{\displaystyle {m \over 2LEk}}
\sin 2\left( ka+\Delta _{o\pm }(k)\right) -%
{\displaystyle {1 \over L}}
{\displaystyle {d\Delta _{o\pm } \over dk}}
\end{equation}

\noindent The charge of the bound state outside the well can be calculated
from Eq. (\ref{bound}) 
\begin{equation}  \label{bch}
Q_{b,ext}=C^2%
{\displaystyle {2m \over \kappa (E+m)}}
e^{-2\kappa a}
\end{equation}

\noindent In the absence of the potential the first two terms in Eqs. (\ref
{d5}, \ref{d7}) would still be there. We wish to calculate $Q_{0,ext}$
defined as the part of $Q_0$ residing outside the well. (Since $Q_0$ itself
vanishes this charge is cancelled exactly by an opposite charge residing
inside the well.) We see that the continuum contribution to the vacuum
charge outside the well resulting from the last term in (\ref{d5}, \ref{d7})
is given by

\begin{equation}
Q_{0,ext}=\frac 1{2\pi }\left( \delta _{+}(\infty )-\delta _{+}(0)-\delta
_{-}(\infty )+\delta _{-}(0)\right)
\end{equation}
\noindent (where the phase shifts $\delta $ refer as before\cite{cdi} to the
sum of the even and odd phase shifts), an expression which is very similar
to Eq. (\ref{err}) and which will in general give non-integral values for $%
Q_{0,ext}.$ To get the total continuum contribution we should integrate over
the terms that depend on the mass explicitly. The final result is in general
non-integral since there is no reason why the charge inside or outside a
particular region of space should be integer-valued.

\medskip\ 

\noindent {\bf Acknowledgements}

\noindent We would like to thank Gabriel Barton, Fred Goldhaber and Mike
Birse for their help and encouragement. One of us (AC) wishes to thank
Professor Volovik and the Low Temperature Laboratory of Helsinki University
of Technology for their hospitality and EU Human Capital and Mobolity
Visitor Programme CHGECT94-0069 for its support.

\end{document}